\documentclass{article}
\usepackage{authblk}
\usepackage[utf8]{inputenc}
\usepackage{diagbox}
\usepackage{amsmath}
\usepackage{amsfonts}
\usepackage{graphicx}
\usepackage{mathtools}
\usepackage{float} 
\usepackage{xcolor}
\usepackage{hyperref}
\usepackage{verbatim}
\usepackage{float}
\providecommand{\keywords}[1]
{
  \small	
  \textbf{\textit{Keywords---}} #1
}
\title{Temporal Consistency Loss for Physics-Informed Neural Networks}
\date{}

\author[1]{Sukirt Thakur}

\affil[1]{\small{School of Mechanical Engineering, 
            Purdue University, 
            West Lafayette,
            47907, 
            Indiana,
            USA}}

\author[2]{Maziar Raissi}

\affil[2]{Department of Applied Mathematics, 
            University of Colorado Boulder, 
            Boulder,
            610101, 
            Colorado,
            USA}

\author[1]{Harsa Mitra}
\author[1]{Arezoo M. Ardekani}

\begin{document}

\maketitle

\begin{abstract}
    Physics-informed neural networks (PINNs) have been widely used to solve partial differential equations in a forward and inverse manner using deep neural networks. However, training these networks can be challenging for multiscale problems. While statistical methods can be employed to scale the regression loss on data, it is generally challenging to scale the loss terms for equations. This paper proposes a method for scaling the mean squared loss terms in the objective function used to train PINNs. Instead of using automatic differentiation to calculate the temporal derivative, we use backward Euler discretization. This provides us with a scaling term for the equations. In this work, we consider the two and three-dimensional Navier-Stokes equations and determine the kinematic viscosity using the spatio-temporal data on the velocity and pressure fields. We first consider numerical datasets to test our method. We test the sensitivity of our method to the time step size, the number of timesteps, noise in the data, and spatial resolution. Finally, we use the velocity field obtained using Particle Image Velocimetry (PIV) experiments to generate a reference pressure field. We then test our framework using the velocity and reference pressure field. \end{abstract}
\keywords{
Physics-informed neural networks, Deep learning, Inverse modelling
}

\section{Introduction}
Physics-informed neural networks \cite{Raissi2019, DHP2018} (PINNs) have become a popular method for solving a wide range of forward and inverse problems. While traditional deep learning methods are data intensive and do not consider the physics of the problem, PINNs leverage the prior information that we have in the form of governing partial differential equations (PDEs). Using the governing equations to regularize the optimization of parameters in PINNs allows us to train large networks with small datasets. This proves handy for problems in biological and engineering systems, as collecting data can be tedious and expensive.

Augmentations to PINNs can be made in five dimensions: 1) More complex physics, 2) more complex geometries, 3) better loss functions, 4) better architectures, and 5) better training processes. While PINNs have been used to solve a whole range of multiphysics problems \cite{Cai2021HT, Kadeethum2020,simnet}, there has been much interest in deploying PINNs to tackle problems in fluid mechanics \cite{Jin2020, Arthurs2021, Cuomo2022, Cai2021}. PINN-based frameworks have been used to model high-speed aerodynamic flows \cite{Mao2020}, porous media flows \cite{Almajid2022}, and biomedical flows \cite{kissas2020}. Recently, PINNs have been used to solve non-Newtonian and complex fluid systems \cite{Mahmoud2022, viscoelasticNet}.

While vanilla feed-forward neural networks remain the most popular architecture, PINNs have been extended to use multiple feed-forward networks \cite{Haghighat2021, Moseley2021}, convolution neural networks \cite{Gao2021, Fang2021}, recurrent neural networks \cite{Zhang2020, Yucesan2021}, and Bayesian neural networks \cite{BPINNs2021}.
However, there are challenges associated with training PINNs. It is not straightforward to train PINNs with ``stiff" PDEs and multiscale problems. There have been numerous efforts to tackle the problem of assigning relative weights to the different objectives. Apart from assigning relative weights through trial and error, the methods include learning rate annealing \cite{Wang2020}, minmax weighting \cite{Liu2021}, using the eigenvalues of the neural tangent kernel matrix \cite{Wang2020NTK} and using the soft self-attention mechanism \cite{McClenny2021}.

In this work, we focus on a better loss function and training process for PINNs. We leverage the scales we have in the observed data to obtain the relative scales of the loss terms. We use backward Euler discretization for time stepping instead of automatic differentiation for the temporal derivative. Other discrete schemes can be used as well, we focus on backward Euler discretization in this work without any loss of generality. This allows us to use the scale in the observed data to scale the governing equations. 

In this work, we consider the two-dimensional and three-dimensional Navier-Stokes equations which govern fluid flows. We obtain the viscosity using the velocity and pressure fields as the observations. We test the sensitivity and robustness of our method to the time step size, the number of timesteps, noise in the data, and spatial resolution for a numerical dataset in section \ref{numDataset}. Here, the number of timesteps refers to the number of discrete time slices we randomly sample from. As our method works robustly, we benchmark our method against the experimental dataset of Particle Image Velocimetry (PIV) observations in section \ref{expDataset}. Finally, we provide some concluding remarks and discuss the future scope of our work in section \ref{conclusion}.

\begin{figure}[!h]
    \centering
    \includegraphics[width=10cm]{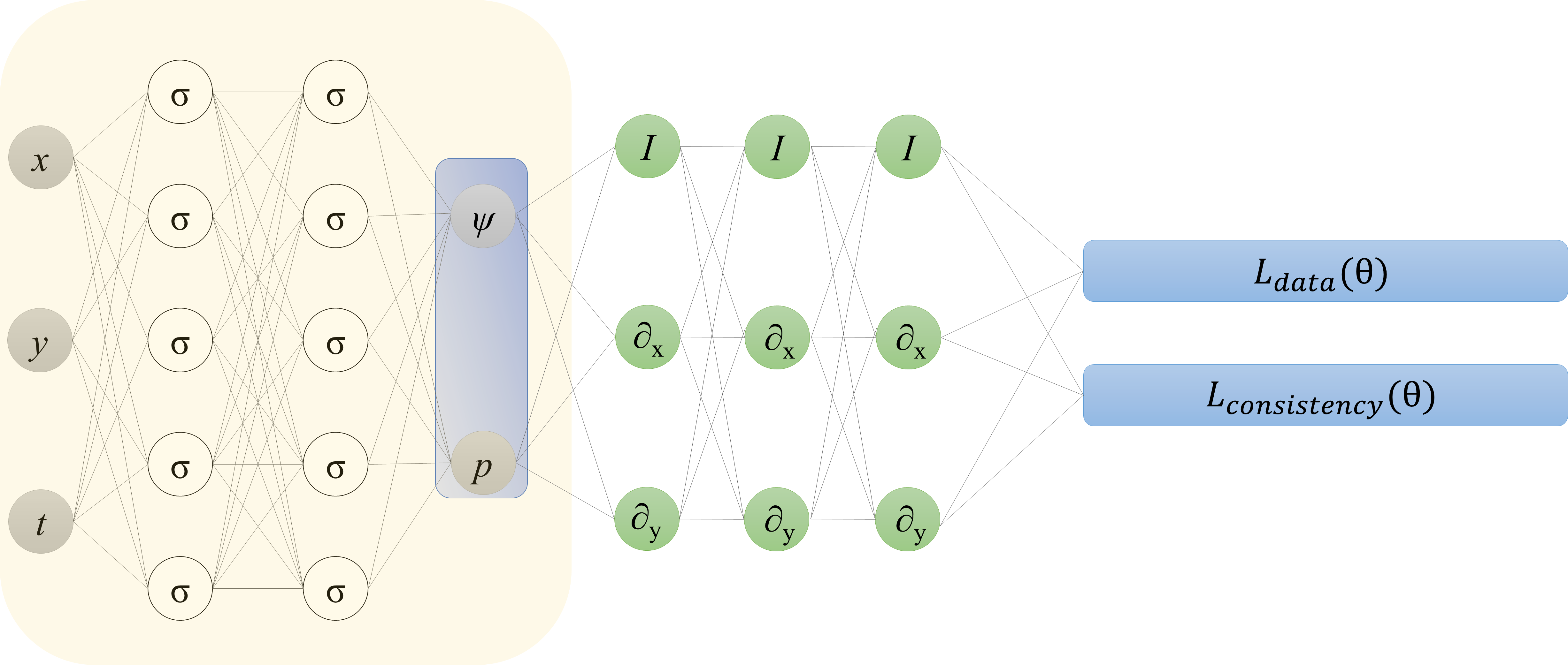}
    \caption{We employ a fully connected neural network with eight hidden layers and 128 neurons per hidden layer to learn the viscosity from velocity and pressure fields in two dimensions. The network takes $t,x,y$ as inputs and outputs the scalar field $\psi$ and the pressure $p$. We employ automatic differentiation to compute the losses described in section \ref{methodology}. Here, $I$ denotes the identity operator, and we compute the differential operators $\partial x$ and $\partial y$ using automatic differentiation. }
    \label{fig:fig_1}
\end{figure}

\section{Methodology}
\label{methodology}
\begin{figure}[!h]
    \centering
    \includegraphics[width=10cm]{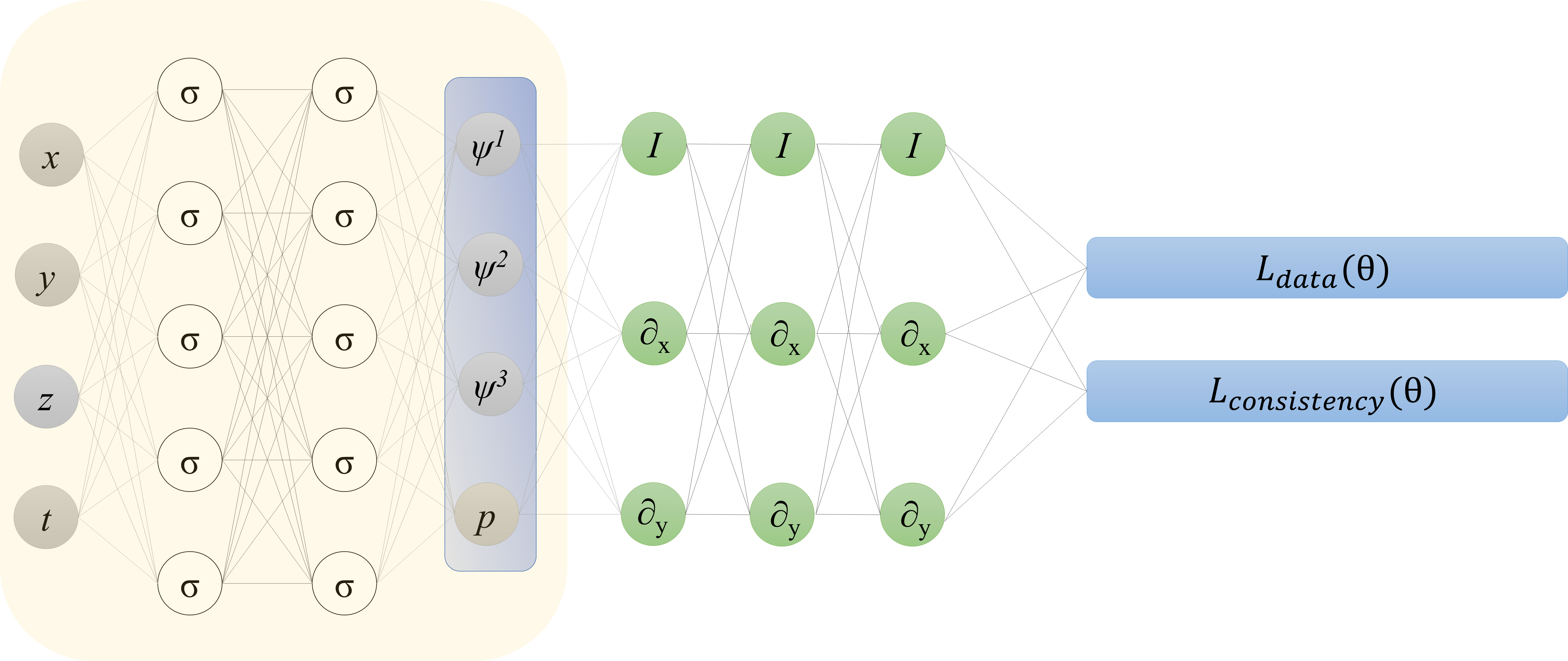}
    \caption{We employ a fully connected neural network with ten hidden layers and 200 neurons per hidden layer to learn the viscosity from velocity and pressure fields in three dimensions. The network takes $t,x,y,z$ as inputs and outputs the components of the vector field $\boldsymbol{\psi}$ and the pressure $p$. We employ automatic differentiation to compute the losses described in section \ref{methodology}. Here, $I$ denotes the identity operator, and we compute the differential operators $\partial x$ and $\partial y$ using automatic differentiation. }
    \label{fig:fig_2}
\end{figure}
\subsection{Fluid Governing Equations}
The conservation of mass for an incompressible fluid is given by 
\begin{equation}\label{continuity}
        \nabla \cdot \boldsymbol{u} = 0,
\end{equation}
where $\boldsymbol{u}$ is the fluid velocity vector. The conservation of momentum of an incompressible Newtonian fluid under isothermal, single-phase, transient conditions in the absence of a body force is given by
\begin{equation}\label{momentum}
     \left( \frac{\partial \boldsymbol{u}}{\partial t} + \boldsymbol{u}\cdot \nabla \boldsymbol{u}\right) = -\frac{1}{\rho}\nabla p + \nu  \nabla^2 \boldsymbol{u},
\end{equation}
where $\rho$ is the density of the fluid, $\boldsymbol{u}$ is the velocity vector, $t$ is the time, $p$ is the pressure, and $\nu$ is the kinematic viscosity. The vector form of the momentum equation in two dimensions in $x$ and $y$ directions is, respectively, given by
\begin{equation}\label{momentum_2D}
\begin{split}
        & u_t + uu_x + vu_y = -\frac{1}{\rho} p_x + \nu (u_{xx} + u_{yy}), \\
        & v_t + uv_x + vv_y = -\frac{1}{\rho} p_y + \nu (v_{xx} + v_{yy}), \\
\end{split}
\end{equation}
where the subscripts denote the derivatives. The momentum equation in vector form in the $x$, $y$ and $z$ directions is, respectively, given by
\begin{equation}\label{momentum_3D}
\begin{split}
        & u_t + uu_x + vu_y + wu_z = -\frac{1}{\rho} p_x + \nu (u_{xx} + u_{yy} + u_{zz}), \\
        & v_t + uv_x + vv_y + wv_z = -\frac{1}{\rho} p_y + \nu (v_{xx} + v_{yy} + v_{zz}), \\
        & w_t + uw_x + vw_y + ww_z = -\frac{1}{\rho} p_z + \nu (w_{xx} + w_{yy} + w_{zz}), \\
\end{split}
\end{equation}
\subsection{Physics informed neural networks}
We define the spatial coordinates in two and three dimensions as
$
    \boldsymbol{x} = (x, y)
$
and 
$
    \boldsymbol{x} = (x, y, z).
$
We define the velocity field of an incompressible isothermal Newtonian fluid as 
\begin{equation}
    \boldsymbol{u}(t,\boldsymbol{x}) = (u(t,\boldsymbol{x}), v(t,\boldsymbol{x})),
\end{equation} in two dimensions and as
\begin{equation}
    \boldsymbol{u}(t,\boldsymbol{x}) = (u(t,\boldsymbol{x}), v(t,\boldsymbol{x}), w(t,\boldsymbol{x})),
\end{equation} in three dimensions. Our observables at the $N$ spatio-temporal data coordinates $\{(t_n, \boldsymbol{x}_n), n = 1, \ldots, N\}$ are the corresponding velocity and pressure fields. We define the velocity field in three dimensions as
\begin{equation}
    \boldsymbol{u} = \nabla \times \boldsymbol{\psi},
\end{equation}
where $\boldsymbol{\psi}$ is a vector in three dimensions. We define the vector $\boldsymbol{\psi}$ with components $\psi^1, \psi^2 $, and $\psi^3$. We get the velocity field as 
\begin{equation}
\begin{split}
    & u = \psi^3_y - \psi^2_z \\
    & v = \psi^1_y - \psi^3_z \\
    & w = \psi^2_y - \psi^1_z,
\end{split}
\end{equation}
where $u, v$ and $w$ are the components of velocity in the $x, y $ and $z$ directions, respectively. By definition, the velocity field will then satisfy the continuity equation \eqref{continuity}. In two dimensions, for the $x$ and $y$ components of velocity, we make the assumption that
\begin{equation}
    u = \psi_y, v = -\psi_x,
\end{equation}
for some latent function $\psi(t,\boldsymbol{x
})$.   We approximate the function $(t,x,y) \longmapsto (\psi, p)$ using a deep neural network with parameters $\theta$ for the two-dimensional case. Here $p$ denotes the pressure field. For the three-dimensional case, a deep neural network with parameters $\theta$ was used to approximate the function $(t,x,y,z) \longmapsto (\psi^1,\psi^2,\psi^3, p)$. The schematic for the neural network setup for the two and three-dimensional cases are shown in fig. \ref{fig:fig_1} and \ref{fig:fig_2}, respectively. We define the mean squared loss for regression over the velocity and pressure fields in two-dimensions as 
\begin{equation}
\begin{split}    
    L_{data}(\theta) = &\mathbb{E}_{(t,x,y,u)}[\frac{|{u}(t,x,y;\theta) - {u}|^2}{{\sigma_u}^2}] + \\             
                    &\mathbb{E}_{(t,x,y,v)}[\frac{|{v}(t,x,y;\theta) - {v}|^2}{{\sigma_v}^2}] + \\
                    &\mathbb{E}_{(t,x,y,p)}[\frac{|{p}(t,x,y;\theta) - {p}|^2}{{\sigma_p}^2}],
\end{split}
\end{equation}
and in three-dimensions as
\begin{equation}
\begin{split}    
    L_{data}(\theta) = &\mathbb{E}_{(t,x,y,z,u)}[\frac{|{u}(t,x,y,z;\theta) - {u}|^2}{{\sigma_u}^2}] + \\             
                    &\mathbb{E}_{(t,x,y,z,v)}[\frac{|{v}(t,x,y,z;\theta) - {v}|^2}{{\sigma_v}^2}] + \\             
                    &\mathbb{E}_{(t,x,y,z,w)}[\frac{|{w}(t,x,y,z;\theta) - {w}|^2}{{\sigma_w}^2}] + \\
                    &\mathbb{E}_{(t,x,y,z,p)}[\frac{|{p}(t,x,y,z;\theta) - {p}|^2}{{\sigma_p}^2}],
\end{split}
\end{equation}
where ${\sigma_u}$, ${\sigma_v}$ and ${\sigma_w}$ are the standard deviation of the $x$, $y$ and $z$ components of the reference velocity field, and ${\sigma_p}$ is the standard deviation of the reference pressure field. Here $\mathbb{E}$ denotes the expectation approximated by the population mean (i.e., mean of the observations $t_n, x_n, y_n, z_n, u_n, v_n, w_n, p_n , n = 1, \ldots, N.$). Now, considering the momentum equation in two-dimensions \eqref{momentum_2D}, we define 
\begin{equation}
    \begin{split}
        g(u,v,p;\nu) = uu_x + vu_y + p_x - \nu(u_{xx} + u_{yy}), \\
        h(u,v,p;\nu) = uv_x + vv_y + p_y - \nu(v_{xx} + v_{yy}),
    \end{split}
\end{equation}
and for three-dimensional momentum equation \eqref{momentum_3D}, we have
\begin{equation}
    \begin{split}
        l(u,v,w,p;\nu) = uu_x + vu_y + wu_z + p_x - \nu(u_{xx} + u_{yy} + u_{zz}), \\
        m(u,v,w,p;\nu) = uv_x + vv_y + wv_z + p_y - \nu(v_{xx} + v_{yy} + v_{zz}), \\
        n(u,v,w,p;\nu) = uw_x + vw_y + ww_z + p_z - \nu(w_{xx} + w_{yy} + w_{zz}).
    \end{split}
\end{equation}
We now create physics-informed neural networks using backward Euler discretization for the time derivative. Other discrete schemes can be used as well, we focus on backward Euler discretization in this work without any loss of generality. For the two-dimensional case, we have
\begin{equation}
    \begin{split}
        u^{pi}(t,x,y;\Delta t, \theta) = u^{pu}(t + \Delta t,x,y; \theta) - \Delta t \,g(&u^{pu}(t + \Delta t,x,y; \theta) \\
        &v^{pu}(t + \Delta t,x,y; \theta) \\
        &p^{pu}(t + \Delta t,x,y; \theta);\nu),
    \end{split}
\end{equation}
\begin{equation}
    \begin{split}
        v^{pi}(t,x,y;\Delta t, \theta) = v^{pu}(t + \Delta t,x,y; \theta) - \Delta t \,h(&u^{pu}(t + \Delta t,x,y; \theta) \\
        &v^{pu}(t + \Delta t,x,y; \theta) \\
        &p^{pu}(t + \Delta t,x,y; \theta);\nu),
    \end{split}
\end{equation}
and the three-dimensional physics-informed neural networks are given by
\begin{equation}
    \begin{split}
        u^{pi}(t,x,y,z;\Delta t, \theta) = u^{pu}(t + \Delta t,x,y,z; \theta) - \Delta t \,l(&u^{pu}(t + \Delta t,x,y,z; \theta) \\
        &v^{pu}(t + \Delta t,x,y,z; \theta) \\
        &w^{pu}(t + \Delta t,x,y,z; \theta) \\
        &p^{pu}(t + \Delta t,x,y,z; \theta);\nu),
    \end{split}
\end{equation}
\begin{equation}
    \begin{split}
        v^{pi}(t,x,y,z;\Delta t, \theta) = v^{pu}(t + \Delta t,x,y,z; \theta) - \Delta t \,m(&u^{pu}(t + \Delta t,x,y,z; \theta) \\
        &v^{pu}(t + \Delta t,x,y,z; \theta) \\
        &w^{pu}(t + \Delta t,x,y,z; \theta) \\
        &p^{pu}(t + \Delta t,x,y,z; \theta);\nu),
    \end{split}
\end{equation}
\begin{equation}
    \begin{split}
        w^{pi}(t,x,y,z;\Delta t, \theta) = w^{pu}(t + \Delta t,x,y,z; \theta) - \Delta t \,n(&u^{pu}(t + \Delta t,x,y,z; \theta) \\
        &v^{pu}(t + \Delta t,x,y,z; \theta) \\
        &w^{pu}(t + \Delta t,x,y,z; \theta) \\
        &p^{pu}(t + \Delta t,x,y,z; \theta);\nu),
    \end{split}
\end{equation}
here the superscript pi denotes a physics-informed network and pu denotes a physics-uninformed network. Since the physics-informed and uninformed networks evaluate the velocities at the same point ${t,x,y}$, they need to be consistent. We enforce this using a consistency loss
\begin{equation}
\begin{split}
    L_{consistency}(\theta;\Delta t) =  & \mathbb{E}_{(t,x,y)}[\frac{|{u}^{pi}(t,x,y;\Delta t,\theta)-{u}^{pu}(t,x,y;\theta)|^2}{{\sigma_{u}}^2}] + \\
    & \mathbb{E}_{(t,x,y)}[\frac{|{v}^{pi}(t,x,y;\Delta t,\theta)-{v}^{pu}(t,x,y;\theta)|^2}{{\sigma_{v}}^2}],
\end{split}
\end{equation}
in two-dimensions. The consistency loss in three-dimensions is defined as
\begin{equation}
\begin{split}
    L_{consistency}(\theta;\Delta t) =  & \mathbb{E}_{(t,x,y,z)}[\frac{|{u}^{pi}(t,x,y,z;\Delta t,\theta)-{u}^{pu}(t,x,y,z;\theta)|^2}{{\sigma_{u}}^2}] + \\
    & \mathbb{E}_{(t,x,y,z)}[\frac{|{v}^{pi}(t,x,y,z;\Delta t,\theta)-{v}^{pu}(t,x,y,z;\theta)|^2}{{\sigma_{v}}^2}] + \\
    & \mathbb{E}_{(t,x,y,z)}[\frac{|{w}^{pi}(t,x,y,z;\Delta t,\theta)-{w}^{pu}(t,x,y,z;\theta)|^2}{{\sigma_{w}}^2}].
\end{split}
\end{equation}
The parameters $\theta$ are then optimized to minimize the following combined loss
\begin{equation}\label{stressLoss}
    L_{MSE}(\theta) =  L_{data}(\theta) + L_{consistency}(\theta).
\end{equation}
\begin{figure}[!h]
    \centering
    \includegraphics[width=\textwidth]{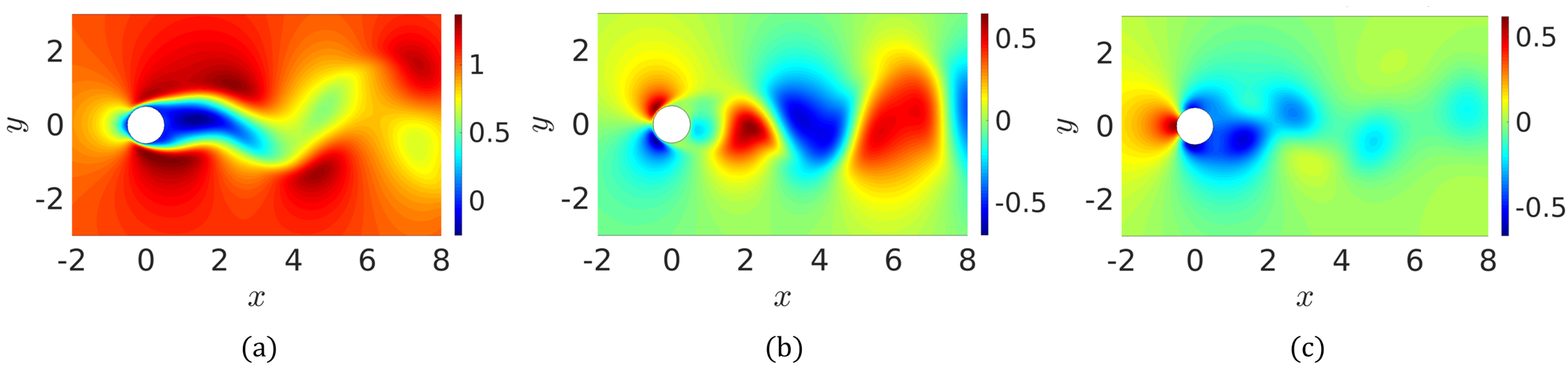}
    \caption{A snapshot of the reference (a) x-velocity, (b) y-velocity and (c) pressure fields of the two-dimensional dataset. }
    \label{fig:fpc}
\end{figure}
\section{Results}
\label{results}
To test our method, we consider two and three-dimensional numerical datasets (section \ref{numDataset}) and an experimental dataset (section \ref{expDataset}). We generated the two-dimensional dataset using the open source CFD toolbox OpenFOAM \cite{Weller198} for the flow past a cylinder. A snapshot of the reference velocity and pressure fields of this dataset is shown in fig. \ref{fig:fpc}. For the three-dimensional case, we look at the flow inside an aneurysm \cite{Raissi2020}. The three-dimensional dataset was generated using the spectral element method, and the dataset is available at https://github.com/maziarraissi/HFM. For the experimental dataset, we calculate the velocity field for water in a channel flow using PIVLab \cite{PIVlab} by tracking particles. A PINN solver was then used to generate the pressure field using the known viscosity of water. We then used the velocity field from PIVlab and the pressure field from the PINN solver to test the method discussed in this paper. 
\begin{figure}[!h]
    \centering
    \includegraphics[height=5.2cm]{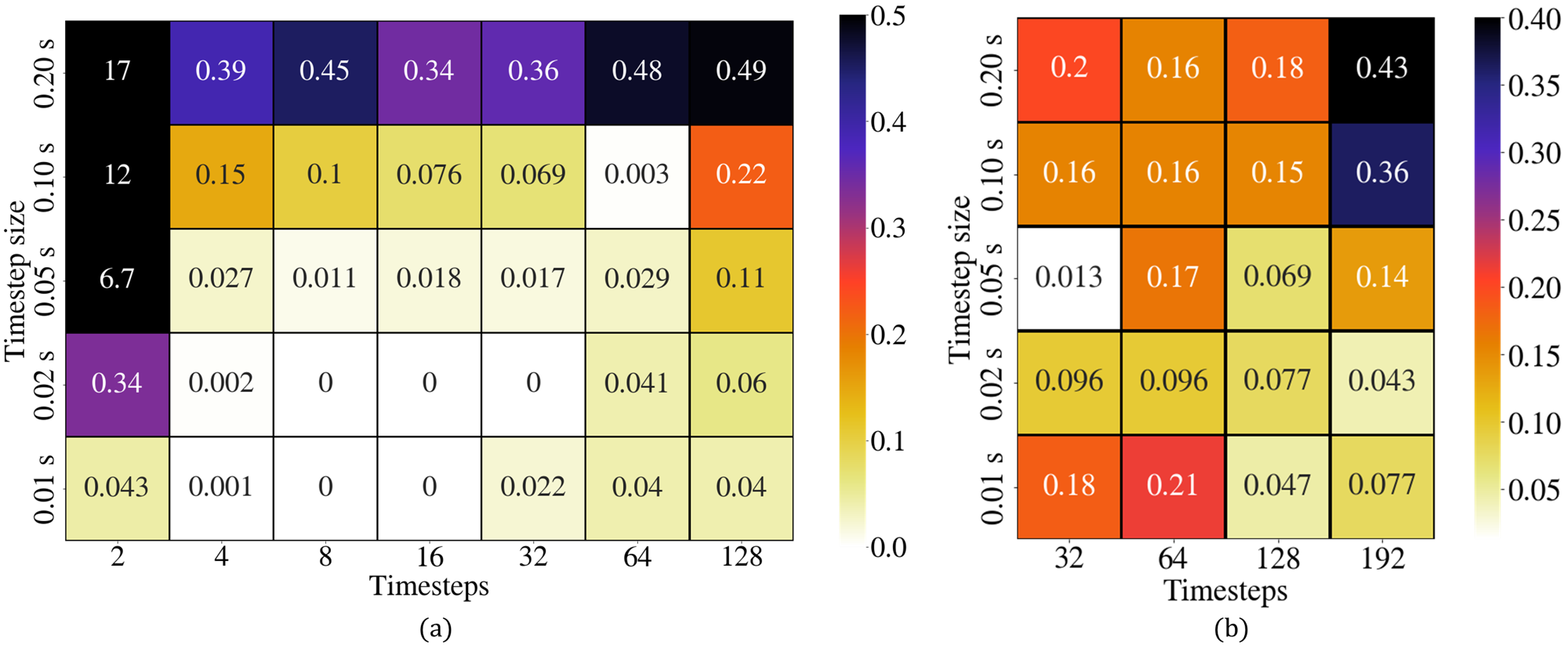}
    \caption{Relative error for viscosity for different combinations of the number of timesteps and timestep size for the (a) two-dimensional and (b) three-dimensional numerical dataset. }
    \label{fig:fig_3}
\end{figure}
\subsection{Numerical datasets}
\label{numDataset}

\begin{figure}[!h]
    \centering
    \includegraphics[height=5.5cm]{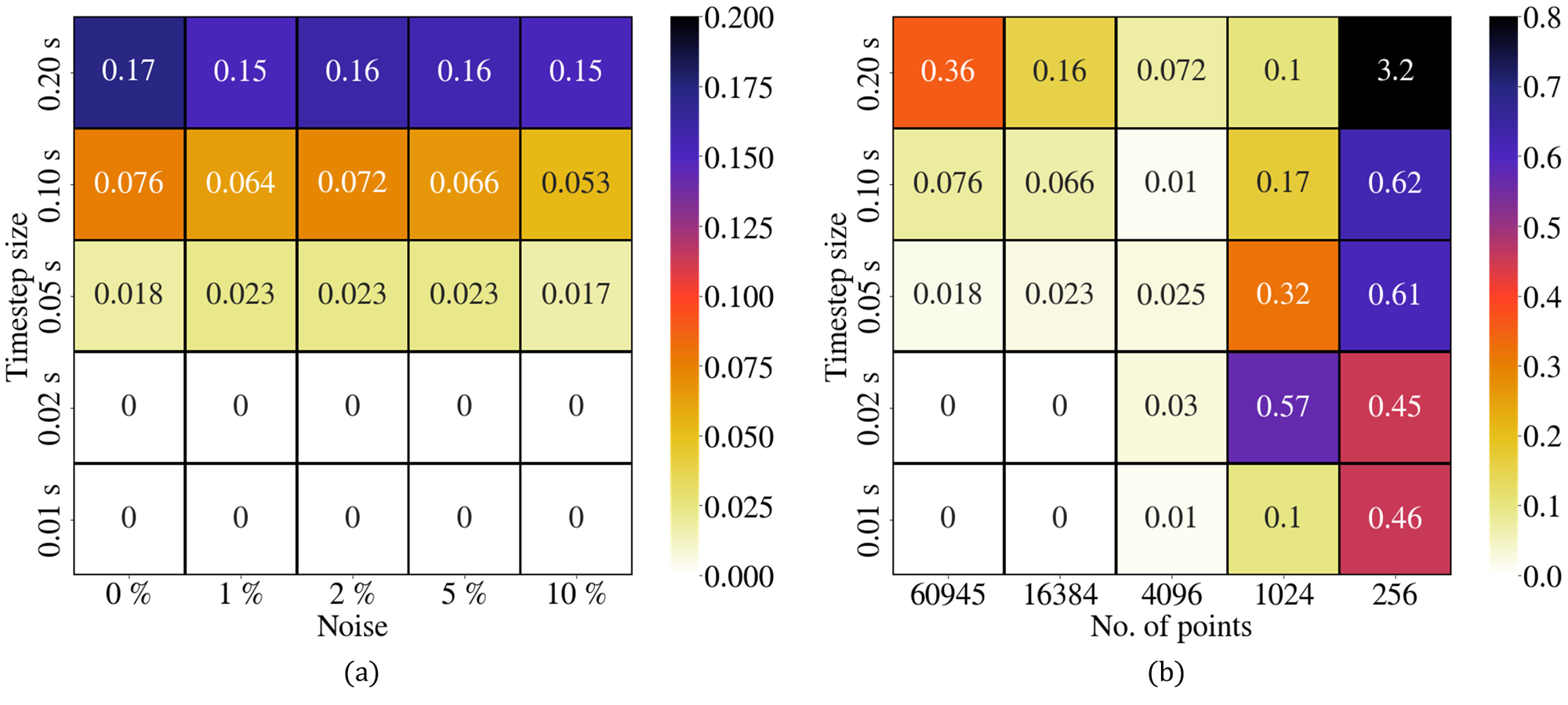}
    \caption{Relative error for viscosity for (a) different combinations of the number of timesteps and noise and (b) different combinations of the number of timesteps and number of spatial points for the two-dimensional numerical dataset. We noticed that the addition of Gaussian noise does not have a significant effect on results.  Our framework eventually breaks down when only 256 points are randomly sampled. }
    \label{fig:fig_4}
\end{figure}
For all the two-dimensional datasets in this section, we present the scalar fields $\psi$ and pressure using an eight-layer deep, fully connected neural network with 128 neurons per hidden layer. For the three-dimensional case, we present $\psi_1, \psi_2, \psi_3$ and pressure using a ten-layer deep neural network with 200 neurons per hidden layer. We use the swish activation function. The use of other architectures might yield better results. A cosine learning rate schedule \cite{Ilya2017} was used in all the runs reported in this work. We used a value of $2.5\text{e-}03$ for $\eta_{max}$ and $2.5\text{e-}06$ for $\eta_{min}$ to get the learning rate $\eta$ as defined in the following equation
\begin{equation}
    \eta =  \eta_{min} + 0.5(\eta_{max} - \eta_{min})\left( 1 + \cos\left( \frac{T_{cur}}{T_{max}}\pi\right)\right),
\end{equation}
where $T_{cur}$ is the current time step and $T_{max}$ is the total timestep. For the two-dimensional case, we choose a mini-batch size of 1024 for the spatio-temporal point cloud inside the domain. The Adam optimizer \cite{Kingma2015} was used to optimize the parameters of the neural network. We ran 100,000 iterations of the Adam optimizer for each two-dimensional case, and every ten iterations of the Adam optimizer took about 0.15 seconds. We used the same learning rate schedule and mini-batch size for the three-dimensional runs. For the three-dimensional cases, we optimized the parameters using 360,000 iterations of the Adam optimizer, where ten iterations took about 0.54 seconds.

\begin{table}[ht]\label{Result_2D}
  \caption{Viscosity for 2D case}
  \begin{center}
    \begin{tabular}{|c|c|c|c|c|c|c|}
      \hline
      \bf \diagbox[width=3.5 cm, height=0.95cm]{Timesteps}{$\Delta T$} &\bf $0.01 s$& \bf $0.02 s$& \bf $0.05 s$ & \bf $0.10 s$ & \bf $0.20 s$\\
      \hline
      2 & $0.01043$ & $0.01335$& $0.07658$& $0.1328$&$0.1795$\\
      \hline
      4 & $0.01001$ & $0.01002$& $0.01027$& $0.01147$&$0.0139$\\
      \hline
      8 & $0.01$ & $0.01$& $0.01011$& $0.01102$&$0.01452$\\
      \hline
      16 & $0.01$ & $0.01$& $0.01018$& $0.01076$&$0.01345$\\
      \hline
      32 & $0.01022$ & $0.01$& $0.01017$& $0.01069$&$0.01358$\\
      \hline
      64 & $0.0104$& $0.01041$& $0.01029$& $0.01093$ &$0.01479$\\
      \hline
      128 & $0.0104$& $0.0106$& $0.01113$& $0.01222$&$0.01494$\\
      \hline
    \end{tabular}
  \end{center}
  \label{ta:timePairs_2D}
\end{table}

\begin{table}[ht]\label{Result_3D}
  \caption{Viscosity for 3D case}
  \begin{center}
    \begin{tabular}{|c|c|c|c|c|c|c|}
      \hline
      \bf \diagbox[width=3.5 cm, height=0.95cm]{Timesteps}{$\Delta T$} &\bf $0.01 s$& \bf $0.02 s$& \bf $0.05 s$ & \bf $0.10 s$ & \bf $0.20 s$\\
      \hline
      32 & $0.0083$ & $0.0092$& $0.01005$& $0.01177$&$0.01222$\\
      \hline
      64 & $0.008$& $0.0092$& $0.01187$& $0.01176$ &$0.01176$\\
      \hline
      128 & $0.0097$& $0.0094$& $0.01088$& $0.01175$ &$0.01204$\\
      \hline
      192 & $0.0094$& $0.009746$& $0.01158$& $0.01388$& $0.01456$\\
      \hline
    \end{tabular}
  \end{center}
  \label{ta:timePairs_3D}
\end{table}

We first tested the sensitivity of our method to the timestep size and the number of time steps. Here, the number of timesteps refers to the number of discrete time slices we randomly sample from. We do this to test the sensitivity of our method to temporal resolution and the amount of data. We report the kinematic viscosity obtained for the two-dimensional dataset in table \ref{ta:timePairs_2D}, where the reference value for the dimensionless kinematic viscosity was $0.01$. For the three-dimensional dataset, the reference dimensionless kinematic viscosity was $0.01018$, and we report the results in table \ref{ta:timePairs_3D}. We show the plot for the relative errors for different combinations of timestep size and timesteps for the two-dimensional and three-dimensional cases in fig. \ref{fig:fig_3}. While the trend is not strictly monotonic, increasing the spatial resolution by decreasing the timestep size and increasing the amount of data improves the results. Our framework reports a low relative error for a wide range of combinations. 

To test the sensitivity of our method to noise, we added Gaussian noise to the two-dimensional dataset. We report the values for viscosity with 16 timesteps with time step sizes $0.01s, 0.02s, 0.05s, 0.10s$, and $0.20s$ at different noise levels in table \ref{ta:noise}. We plot the relative errors for different combinations of timestep sizes and Gaussian noise in fig. \ref{fig:fig_4}. We observed that the amount of Gaussian noise did not significantly affect the error, and our method worked well even when 10\% Gaussian noise was added to the dataset. This result was in agreement with what was observed for PINNs earlier \cite{viscoelasticNet}.

The low sensitivity to Gaussian noise might result from many spatial points in the dataset. We trained our model on fewer spatial points to test this. We randomly sampled 60495, 16384, 4096, 1024, and 256 spatial points at 16 time steps and added $5\%$ Gaussian noise. We report the predicted viscosities for each case in table \ref{ta:spatialPts} and show the relative error in fig. \ref{fig:fig_4}. We obtained good results by randomly sampling 4096 points, or roughly 1 in 16 spatial points. Our method worked well for smaller time step sizes and eventually broke down when we randomly sampled only 256 spatial points or around 1 in 256 spatial points. Since our setup worked for sparse and noisy data, we next considered a real-world dataset obtained through experiments.

\begin{table}[ht]\label{Noise_2D}
  \caption{Viscosity for 2D case as a function of noise level for 16 timesteps with 5\% noise}
  \begin{center}
    \begin{tabular}{|c|c|c|c|c|c|c|}
      \hline
      \bf \diagbox[width=3.5 cm, height=0.95cm]{Noise}{$\Delta T$} &\bf $0.01 s$& \bf $0.02 s$& \bf $0.05 s$ & \bf $0.10 s$ & \bf $0.20 s$\\
      \hline
      0\% noise & $0.01$ & $0.01$& $0.01018$& $0.01076$&$0.01173$\\
      \hline
      1\% noise & $0.01 $& $ 0.01$& $0.01023 $& $0.01064 $ &$ 0.01152$\\
      \hline
      2\% noise & $0.01$& $0.01$& $0.01023$& $0.01072$ &$0.0116 $\\
      \hline
      5\% noise & $0.01$& $0.01 $& $0.01023 $& $ 0.01066$& $0.01156 $\\
      \hline
      10\% noise & $ 0.01$& $0.01 $& $0.01017 $& $ 0.01053$& $ 0.01151$\\
      \hline

    \end{tabular}
  \end{center}
  \label{ta:noise}
\end{table}

\begin{table}[ht]\label{Spatial_2D}
  \caption{Viscosity for 2D case as a function of spatial points for 16 timesteps}
  \begin{center}
    \begin{tabular}{|c|c|c|c|c|c|c|}
      \hline
      \bf \diagbox[width=3.5 cm, height=0.95cm]{\# points}{$\Delta T$} &\bf $0.01 s$& \bf $0.02 s$& \bf $0.05 s$ & \bf $0.10 s$ & \bf $0.20 s$\\
      \hline
      60945 & $0.01$ & $0.01$& $0.01018$& $0.01076$&$ 0.01357 $\\
      \hline
      16384 & $0.01$& $0.01 $& $0.01023 $& $ 0.01066$& $0.01156 $\\
      \hline
      4096 & $ 0.0099$& $0.0097$& $0.01025$& $0.0101$&$ 0.01072 $\\
      \hline
      1024 & $0.011$ & $0.0157$& $0.0132$& $0.0117$& $ 0.01107$\\
      \hline
      256 & $0.05447$ & $0.05492$& $0.03893$& $0.03758$& $ 0.04164$\\
      \hline
    \end{tabular}
  \end{center}
  \label{ta:spatialPts}
\end{table}

\subsection{Experimental Dataset}
\label{expDataset}

\begin{figure}[H]
    \centering
    \includegraphics[height=7cm]{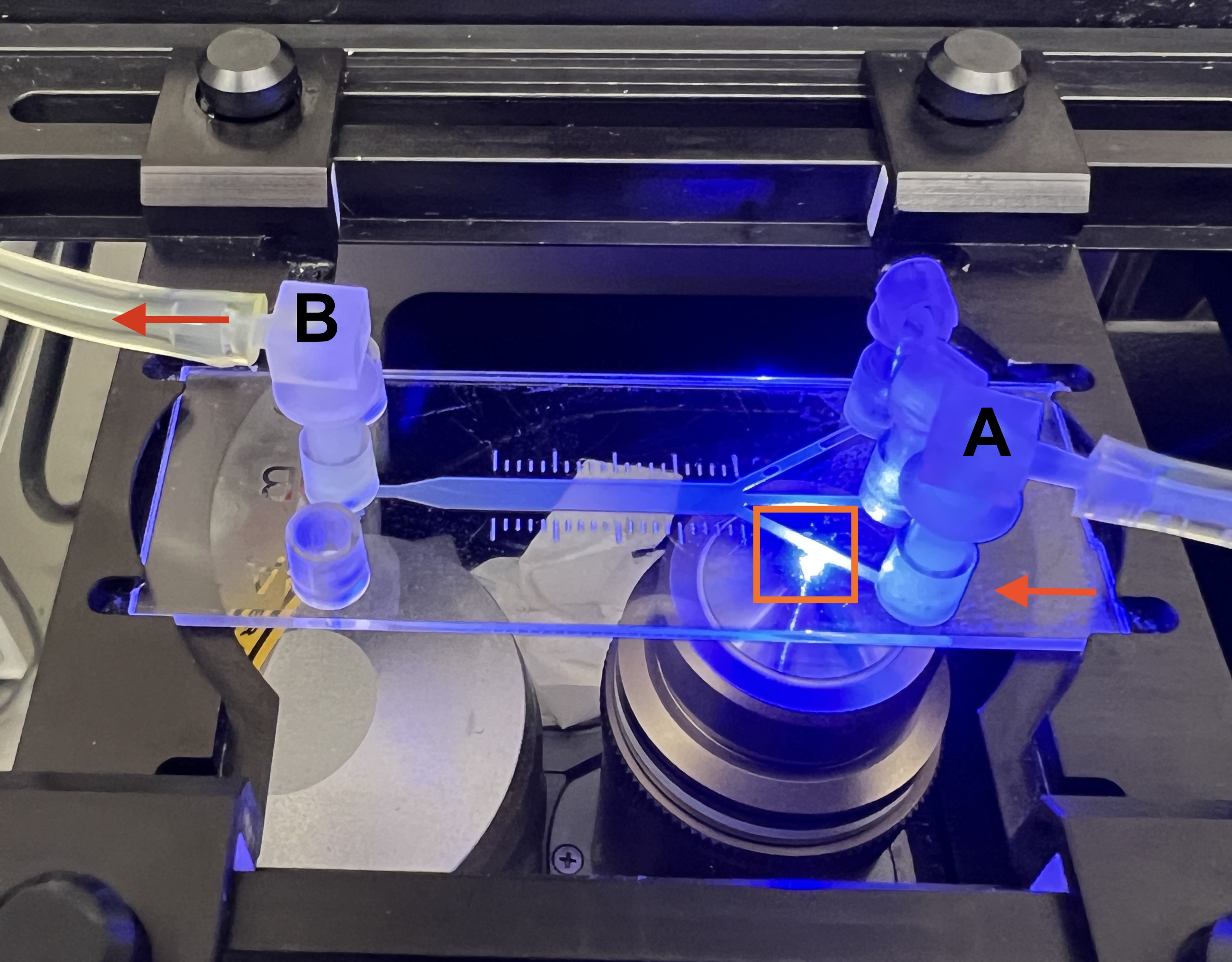}
    \caption{The experimental setup with the µ-Slide III 3in1 (ibidi Inc., WI, USA) is used for the PIV measurements. The flow inlet and outlet are labeled as A and B, respectively. The interrogation area (not to scale) is also represented using the orange square. During the experiment, the other two inlets were closed using ibidi luer locks.}
    \label{fig:expt_setup}
\end{figure}

Water seeded with 1 $\mu$m fluorescent polystyrene beads (Bangs Laboratories Inc., IN, USA) at 2\% (w/w) concentration is used for the experimental validation. As shown in Fig.~\ref{fig:expt_setup}, a syringe pump drives the fluid flow within an oblique channel of 1 mm width and 0.4 mm height. We applied water flow at 40 $\mu$l/min. A 520 nm laser using an inverted microscope coupled with a confocal system (Nikon, NY, USA) for imaging and used an oil immersion 60x (0.1083 $\mu$m/px) lens. We collected 3000 images at 5 ms intervals (200 fps). \par

For post-processing PIVlab MATLAB GUI is used~\cite{PIVlab}. We imported the images in the pairwise sequencing scheme. Image pre-processing using the PIVlab interface is also applied to remove the background light intensity. The 2-D velocity field is extracted in the $x$-$y$ plane using the Fast Fourier Transform (FFT) window deformation algorithm, along with three passes, i.e., 128, 64, and 32 pixel interrogation areas. Finally, the mean- $x$ and $y$ velocity components are calculated and exported separately. 

We use the velocity field obtained from PIVLab to generate a reference pressure field. Our framework then uses the velocity and pressure fields to predict water viscosity at room temperature. We used an eight-layer deep, fully connected neural network with 128 neurons per hidden layer. We used the learning schedule described in section \ref{numDataset}, and the parameters of the network were optimized using 800,000 iterations of the Adam optimizer. The reference value for the water viscosity at room temperature is 0.01 poise \cite{Swindells1969}, and the value we get from our model is 0.00977 poise. 

\section{Conclusions and Future Work}
\label{conclusion}

It is generally challenging to assign relative weights to the loss terms while training physics-informed neural networks, especially with multiscale data. We propose a novel solution for this challenge. By using backward Euler discretization for temporal derivatives instead of automatic differentiation, we can use the data's statistical properties to get the loss terms' relative weights. In this work, we consider the two and three-dimensional Navier-Stokes equations and determine the kinematic viscosity using spatio-temporal data on the velocity and pressure fields.

For the two-dimensional case, we look at the flow past a cylinder and flow in an aneurysm for the three-dimensional case. We test the sensitivity and robustness of our method against the timestep size, the number of timesteps, noise in the data, and the spatial data resolution. Since our method worked well for a wide range of numerical data, we tested our method using experimental data. We used the velocity field from experimental PIV measurements of a channel flow to generate a reference pressure field. We tested our framework using this velocity and reference pressure fields to get water viscosity at room temperature. We demonstrated that our framework worked well with an experimental dataset. This work uses spatio-temporal data on the pressure and velocity fields as input. For future work, using just the velocity field as an input and solving for the pressure field can be explored. Then the velocity field from PIV measurements can be used directly to learn the viscosity and the pressure field for both two and three-dimensional flows. 

\section{Acknowledgements}

A.M.A. acknowledges financial support from the National Science Foundation (NSF) through  Grant No.  CBET-2141404.

\bibliographystyle{elsarticle-num}
\bibliography{references}

\end{document}